# Superfluid density of superconductor-ferromagnet bilayers

Thomas R. Lemberger,[*] Iulian Hetel, Adam J. Hauser, and F.Y. Yang
Department of Physics, The Ohio State University
191 W. Woodruff Ave., Columbus, OH 43210-1117

We report the first measurements of the effective superfluid density $n_S(T) \propto \lambda^{-2}(T)$ of Superconductor-Ferromagnet (SC/FM) bilayers, where $\lambda$ is the effective magnetic field penetration depth. Thin Nb/Ni bilayers were sputtered in ultrahigh vacuum in quick succession onto oxidized Si substrates. Nb layers are 102 Å thick for all samples, while Ni thicknesses vary from 0 to 100 Å. $T_C$ determined from $\lambda^{-2}(T)$ decreases rapidly as Ni thickness $d_{Ni}$ increases from zero to 15 Å, then it has a shallow minimum at $d_{Ni} \approx 25$ Å. $\lambda^{-2}(0)$ behaves similarly, but has a minimum several times deeper. In fact, $\lambda^{-2}(0)$ continues to increase with increasing Ni thickness long after $T_C$ has stopped changing. We argue that this indicates a substantial superfluid density inside the ferromagnetic Ni films.

I. INTRODUCTION

1. Background

The proximity effect between superconductors (SC) and nonsuperconductors has been of interest for many decades[1] due both to fundamental physics associated with interfaces and technical interest in making electrical contact to superconductors for use in devices or high-field magnets. Essentially, when a superconductor makes good contact with a nonsuperconducting material, Cooper pairs diffuse across the interface, lowering the superconducting transition temperature $T_C$ and altering the superconducting density of states. Recently, interest has focused on superconductor/ferromagnet (SC/FM) bilayers, trilayers, etc., because of the theoretical prediction that the superconducting order parameter $\Psi(\mathbf{r})$ oscillates inside the FM layer. The oscillations lead to novel counterintuitive properties not seen in the proximity effect between superconductors and nonmagnetic materials.

According to theory, the order parameter decay length and the oscillation length should be about the same because both are determined by the FM exchange field.[2-7] The oscillation of $\Psi$ leads to the striking prediction that the superconducting transition temperature, $T_C$, decreases but also oscillates as the thickness of the FM film increases. Starting with Jiang et al.,[8] many groups have observed a minimum in $T_C$ via resistance measurements of SC/FM bilayers and multilayers. The oscillation of $\Psi$ was first observed by Ryazanov et al.[9] through oscillations in the critical current of SC/FM/SC Josephson weak links ("pi-junctions"). For a review of experimental work, see Buzdin et al.[7]

In the present work, we probe superconductivity in SC/FM proximity effect bilayers by measuring their ability to screen the magnetic field produced by a small coil. We obtain both $T_C$ and insight into the superfluid density inside the ferromagnetic film. At our low experimental frequencies, the screening factor is proportional to the effective density of superconducting electrons $n_S(T) \propto \lambda^{-2}(T)$, where $\lambda$ is defined below. Consistent with previous work on Nb/Ni bilayers,[10] we find that $T_C$ has a shallow minimum as a function of $d_{Ni}$. $\lambda^{-2}(0)$ has a deeper minimum.

2. Theory

Our experiment is done in the low-frequency, thin-film limit. In this limit, the <u>areal</u> superfluid density of a bilayer is proportional to the imaginary part $Y_2(T)$ of the sheet conductivity $Y(T)$ of the bilayer. In effect, $Y_2$ is the local conductivity $\sigma_2(\omega,z,T)$ integrated through the bilayer:

$$Y_2(\omega,T) = \int_{-d_S}^{d_F} dz\, \sigma_2(\omega,z,T) \approx d_S \sigma_{2S} + d_F \sigma_{2F}, \qquad (1)$$

where the z-axis is perpendicular to the bilayer, and the origin is at the SC-FM interface. We define the effective penetration depth, $\lambda$, from: $\lambda^{-2} \equiv \mu_0 \omega Y_2 / d_S$, dividing by the superconducting film thickness only. Thus, $\lambda^{-2}$ is the superfluid density that the Nb film would have to have in order to produce the same magnetic field screening that a bilayer produces.

Theory of the effective superfluid density of SC/FM bilayers has not yet been developed. We employ a phenomenological analysis. First, we use pairbreaking theory[11] to estimate the superfluid density in the Nb layers. Subtracting this from our data, we obtain a reasonable estimate of the superfluid density in the Ni films.

Dirty-limit theory of superconductivity[12] provides results for the magnitude and T-dependence of $1/\lambda^2$ in a homogeneous superconductor:

$$\frac{\lambda^{-2}(T)}{\lambda^{-2}(0)} = \frac{\Delta_0(T)}{\Delta_0(0)} \tanh\left[\frac{\Delta_0(T)}{2k_B T}\right], \qquad (2)$$

where the normalized energy gap, $\Delta_0(T)/\Delta_0(0)$, is well approximated as:[13]

$$\frac{\Delta_0(T)}{\Delta_0(0)} \approx \left[\cos\left(\frac{\pi T^2}{2T_{C0}^2}\right)\right]^{1/2}. \qquad (3)$$

The subscript "0" denotes that these results apply to Nb films alone. In the dirty limit, the elastic scattering rate is large, $\hbar/\tau \gg \Delta(0)$, and the magnitude of $\lambda^{-2}(0)$ is:

$$\lambda^{-2}(0)\big|_{BCS} = \frac{\pi\mu_0\Delta_0(0)}{\hbar\rho_0}, \qquad (4)$$

where $\mu_0$ is the permeability of vacuum and $\rho_0$ is the residual resistivity. The factor of $\Delta_0$ in Eq. (4) indicates that $1/\lambda^2(0)$ of the Nb films would decrease in proportion to $T_C$ if BCS dirty limit theory applied to our bilayers.

If Ni suppresses superconductivity in Nb *via* a "pairbreaking" mechanism,[10] there is an additional decrease in $\lambda^{-2}$ beyond simple BCS theory. There is a pairbreaking energy, $\hbar/\tau_{pb}$, that can be estimated from the suppression of $T_C$:[11]

$$\frac{\pi\hbar}{4\tau_{pb}} \approx k_B(T_{C0} - T_C). \qquad (5)$$

The order parameter, $\Delta(0)$, is reduced below its unperturbed value:

$$\Delta(0) \approx \Delta_0(0) - \frac{\pi\hbar}{4\tau_{pb}}. \qquad (6)$$

Finally, $\lambda^{-2}(0)$ in the Nb film is suppressed in proportion to $\hbar/\tau_{pb}$:[11]

$$\frac{\lambda^{-2}(0)}{\lambda^{-2}(0)\big|_{BCS}} \approx 1 - 0.42\frac{\hbar}{\tau_{pb}\Delta(0)} \approx \frac{0.27 + 1.53 T_C/T_{C0}}{0.8 + T_C/T_{C0}}. \quad [\hbar/\tau_{pb}\Delta(0) \leq 1] \qquad (7)$$

In Eq. (7), $\lambda^{-2}(0)|_{BCS}$ is calculated from Eq. (4) using the measured $T_C$ of the Nb/Ni bilayers. For samples of the most interest, $T_C/T_{C0} \approx 2.9/7.5$, and this factor is about 0.7.

As for the T-dependence of $\lambda^{-2}(T)$, because pairbreaking smears out the peak in the superconducting density of states at $E = \Delta$ and lowers the gap edge, its effect can be approximated by reducing the gap ratio, $\Delta(0)/kT_C$ below the BCS weak coupling value of 1.765. In the limit of strong pairbreaking, $\lambda^{-2}(T)$ decreases as $T^2$ at low T.[11] We will use this approximation to estimate $\lambda^{-2}(0)$.

## II. EXPERIMENTAL DETAILS

Nb/Ni bilayers are deposited by DC sputtering from 2" diameter Nb and Ni targets onto oxidized Si substrates located 2" above the target. Si substrates nominally 18×18×0.4 mm$^3$ are placed into a load-locked UHV chamber with a background pressure of $5\times10^{-10}$ torr. In rapid succession, Nb(102 Å)/Ni($d_{Ni}$) bilayers are deposited followed by a protective layer of Ge(200 Å) with $d_{Ni}$ varying from 0 to 100 Å. The deposition rates are 1.5, 0.94, and 2.0 Å/s for Nb, Ni, and Ge, respectively. The Ge layer is to inhibit oxidation in air. Substrates are nominally at room temperature during deposition. Film thicknesses are calibrated by growing thick films, measuring their thicknesses with an atomic force microscope, and thereafter assuming that film thicknesses are proportional to sputtering time. Calibrations are checked regularly. The UHV environment and precise sample control provide high reproducibility of sample properties. The variation of $T_C$ and superfluid density from run to run is within about 2%, which makes this study possible with sufficient reliability.

We measure at 50 kHz the mutual inductance of coaxial coils located on opposite sides of the bilayer samples. The coils are solenoids nominally 2 mm in diameter and 2 mm in length, much smaller than the areal dimensions of the bilayers. Induced currents lie in the plane of the sample so that the conductivities of Nb and Ni films are measured in parallel; no current crosses the interface between them. Various steps involved in converting mutual inductance to sheet conductivity, Y, have been described.[14,15]

Our results are presented in terms of an effective magnetic penetration depth, $\lambda(T)$, defined from the measured $Y_2$:

$$\lambda^{-2} \equiv \mu_0 \omega Y_2 / d_S, \tag{8}$$

where $d_S = 102$ Å is the Nb film thickness. Because we divide by $d_S$ in the definition of $\lambda^{-2}$ in Eq. (8), $\lambda^{-2}$ represents the equivalent superfluid density of a Nb film that screens magnetic field the same as the bilayer. We believe that our experimental method determines $Y_2$ to < 3% accuracy.

Precautions are taken to ensure that data are taken in the linear-response regime, where the data are independent of the size of the 50 kHz current in the drive coil. We estimate that the Ginzburg-Landau coherence length, $\xi(0)$, is comparable to the Nb film thickness, so the order parameter should have some variation through the Nb films. This is not taken into account in our analysis.

### III. EXPERIMENTAL RESULTS AND DISCUSSION

Figure 1 shows $\lambda^{-2}(T)$ vs. T for several representative Nb(102 Å)/Ni($d_{Ni}$) bilayers. Parameters for all the bilayers that we measured are in Table I. Note that a Nb(102 Å) film alone would have $T_C \approx 7.5$ K and $\lambda^{-2}(0) \approx 34$ $\mu m^{-2}$.[16] Clearly, $T_C$ and $\lambda^{-2}(0)$ for the bilayers with Ni thicknesses of 15 Å or more are greatly suppressed. To estimate $\lambda^{-2}(0)$ for the bilayers, we fitted the lowest temperature data to a two-parameter quadratic: $A - CT^2$, which strictly applies in the strong-pairbreaking limit [11] To get $T_C$, we fitted data not too close to $T_C$ to a three parameter quadratic: $-B(T_C - T) - C(T_C - T)^2$. These fits are indicated by dashed curves in Fig. 1. The abrupt drop in $\lambda^{-2}(T)$ that consistently appears as T approaches $T_C$ is likely due to inhomogeneity of unknown origin.

The dependences of $T_C$ and $\lambda^{-2}(0)$ on $d_{Ni}$ are shown in Fig. 2. Note that $T_C$ and $\lambda^{-2}(0)$ decrease rapidly up to about 20 Å Ni because ferromagnetic moments in Ni suppress superconductivity in the adjacent Nb layers. After $d_{Ni}$ = 20 Å, both $T_C$ and $\lambda^{-2}(0)$ stop decreasing and reverse the trend. $T_C$ increases about 8% while $\lambda^{-2}$ increases about 45%. Above about 100 Å of Ni, properties stop changing.

A microscopic theory is needed. A rough analysis indicates that the superfluid density inside the Ni film is significant, as follows. The superfluid density inside the Nb layers is estimated to be: 34 $\mu m^{-2}$ × ($T_C/T_{C0}$) × 0.7, the latter factor due to pairbreaking [Eq. (7)]. Consider the Nb/Ni bilayer with 25 Å of Ni, which has an effective superfluid density $\lambda^{-2}(0)$ = 8.3 $\mu m^{-2}$. Within uncertainties, our simple model predicts about this value for the Nb film alone. Note that if we had not included pairbreaking, our model could not have accounted for such a low superfluid density. We cannot determine the superfluid density in the Ni film of this bilayer directly. However, most of the 3.4 $\mu m^{-2}$ (40%) increase in the effective superfluid density of the bilayers as Ni thickness increases to 40 Å has to be assigned to the Ni film, since $T_C$ increases by only 7%. Let us assume that the superfluid density of the Nb film increased by 7%, i.e., by about 0.6 $\mu m^{-2}$. The Ni film has to account for the rest. Because the Ni film is 2.5 times thinner than the Nb film, its volume superfluid density would have to be: $\lambda^{-2}(0) \approx 2.5 \times 2.8$ $\mu m^{-2} \approx 7$ $\mu m^{-2}$, comparable to the Nb film, to account for the observed increase in screening between coils.

## IV. CONCLUSION

We measured the effective superfluid densities $\lambda^{-2}(T) \propto n_S(T)$ of a series of Nb/Ni bilayers with Nb thickness 102 Å. $\lambda^{-2}(0)$ shows a pronounced minimum at a Ni thickness of about 25 Å. A phenomenological analysis assigns the increase of $\lambda^{-2}(0)$ with increasing Ni thickness to a substantial superfluid density inside the ferromagnetic Ni films.

## ACKNOWLEDGEMENTS

I. H. gratefully acknowledges support of an OSU Presidential Fellowship. We are grateful to Profs. Julia Meyer, A. Buzdin, and A. Golubov for many useful discussions and to Michael Hinton for technical assistance.

TABLE I.

| $d_{Ni}$ (Å) | $T_C$ (K) (±30 mK) | $\lambda^{-2}(0)$ (μm$^{-2}$) (±5%) |
|---|---|---|
| 0 | 7.5 | 34 |
| 15 | 3.14 | 9.0 |
| 20 | 2.89 | 8.3 |
| 25 | 2.72 | 8.3 |
| 30 | 2.74 | 8.8 |
| 35 | 2.71 | 9.4 |
| 37.5 | 2.96 | 11.5 |
| 40 | 2.95 | 11.7 |
| 50 | 2.94 | 11.5 |
| 75 | 3.02 | 13.8 |
| 100 | 2.83 | 12.0 |

TABLE I. Parameters for Nb/Ni bilayers with $d_{Nb} = 102$ Å. Uncertainties in $T_C$ and $\lambda^{-2}(0)$ are dominated by uncertainties in extrapolations $\lambda^{-2}(T) \to 0$ and $T \to 0$, respectively.

FIGURE CAPTIONS

FIG. 1. $\lambda^{-2}(T)$ vs. T for representative Nb/Ni bilayers with $d_{Nb}$ = 102 Å. Dashed curves are extrapolations to T = 0 and T = $T_C$ explained in the text.

FIG. 2. $T_C$ (squares) and $\lambda^{-2}(0)$ (circles) vs. $d_{Ni}$ for Nb/Ni bilayers with $d_{Nb}$ = 102 Å

Figure 1.

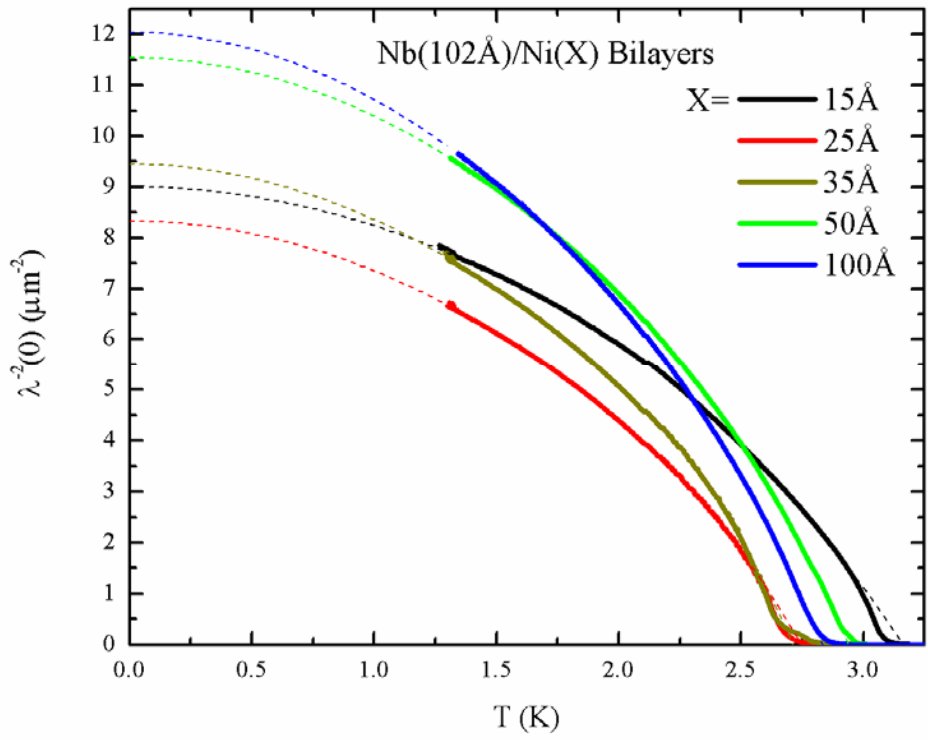

FIG. 1. $\lambda^{-2}(T)$ vs. T for representative Nb/Ni bilayers with $d_{Nb}$ = 102 Å.
Dashed curves are extrapolations to T = 0 and T = $T_C$ explained in the text.

Figure 2.

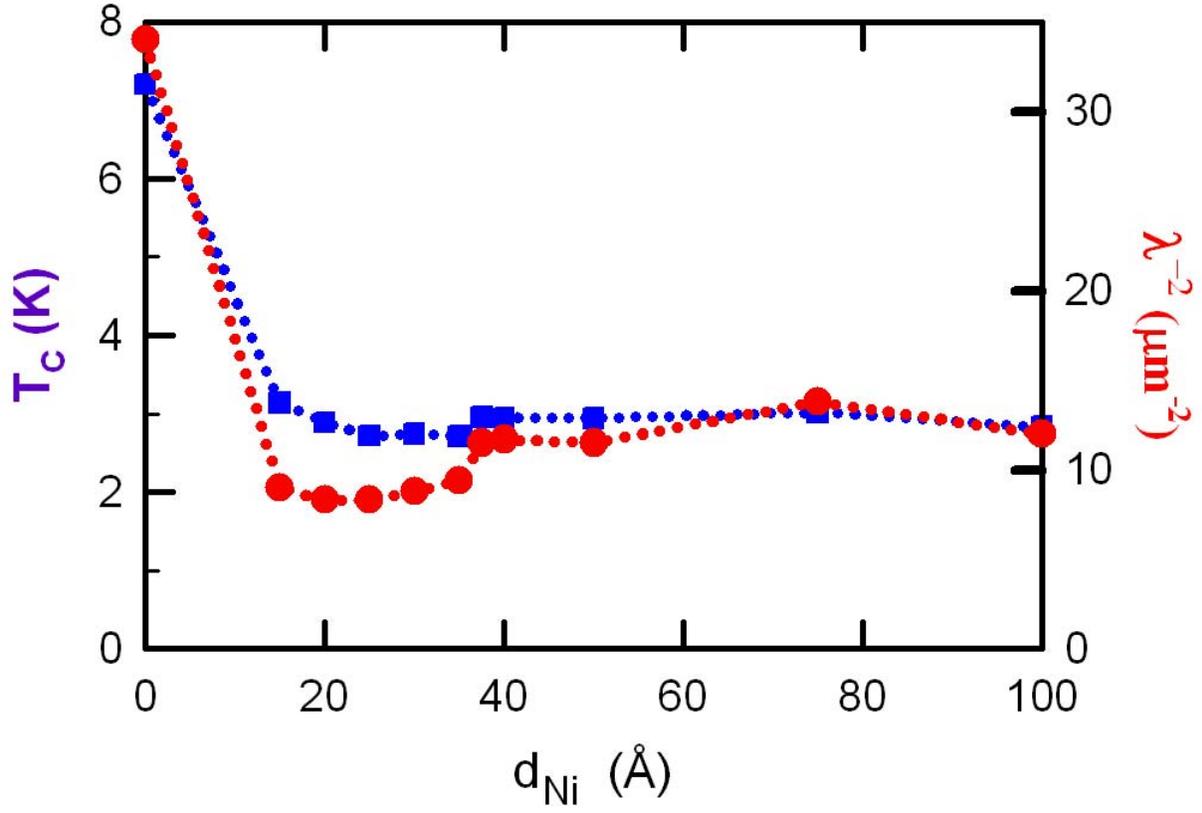

FIG. 2. $T_C$ (squares) and $\lambda^{-2}(0)$ (circles) vs. $d_{Ni}$ for Nb/Ni bilayers with $d_{Nb} = 102$ Å.